\documentclass[cits]{PoS}
\usepackage{amsmath,amssymb}
\usepackage{units}
\usepackage{setspace}
\newcommand{\preprint}{
  \begin{picture}(0,0)
    \put(0,110){{\rm\normalsize HU-EP-10/12}}
  \end{picture}}

\title{\preprint
       QCD Lambda parameter from Landau-gauge gluon and ghost correlations}

\ShortTitle{QCD Lambda parameter from Landau-gauge gluon and ghost correlations}
\author{\speaker{A.~Sternbeck}$^{\,ab}$, E.--M.~Ilgenfritz${}^{c}$,
  K.~Maltman$^{ad}$, M.~M{\"u}ller-Preussker${}^c$,
  L.~von~Smekal$^{ae}$ and A.~G.~Williams$^a$\\\\
  \llap{$^a$} CSSM, School of Chemistry and Physics, University of
  Adelaide, SA 5005, Australia\\
  \llap{$^b$} Institut f\"ur Theoretische Physik, Universit\"at
  Regensburg, D-93040 Regensburg, Germany\\
  \llap{$^c$} Humboldt-Universit\"at zu Berlin, Institut f\"ur Physik,
  D-12489 Berlin, Germany\\
  \llap{$^d$} Department of Mathematics and Statistics, York Univ.,
  Toronto, ON, M3J 1P3, Canada\\
  \llap{$^e$} Institut f\"ur Kernphysik, Technische Universit\"at
  Darmstadt, D-64289 Darmstadt, Germany}

\renewcommand{\Re}{\operatorname{\mathfrak{Re}}}      
\newcommand{\Tr}{\operatorname{Tr}}                   
\newcommand{\MS}{\overline{\mathsf{MS}}}
\newcommand{\MOM}{\mathsf{MOM}}
\newcommand{\MSb}{\MS}
\newcommand{\MM}{\mathsf{MM}}
\newcommand{\alphaMS}{\alpha^{\MS}_s}
\newcommand{\alphaMSref}{\alpha^{\MSb}_{\mathrm{ref}}}
\newcommand{\LambdaMS}{\Lambda_{\MSb}}
\newcommand{\LambdaMM}{\Lambda_{\MM}}
\newcommand{\alphaMM}{\alpha^{\MM}_s}
\newcommand{\alphaMML}{\alpha^{\MM}_L}
\newcommand{\RS}{\mathtt{S}}

\newcommand{\Eq}[1]{{Eq.~\eqref{#1}}}
\newcommand{\Fig}[1]{Fig.~\ref{#1}}
\newcommand{\Tab}[1]{Tab.~\ref{#1}}

\abstract{We utilise a recently developed minimal MOM scheme
  to determine the QCD Lambda parameter from the
  gluon and ghost propagators in lattice Landau gauge. We 
  discuss uncertainties in the analysis and report our preliminary 
  zero and two flavour results, which are $r_0\LambdaMS^{(0)}=0.62(1)$ and
  $r_0\LambdaMS^{(2)}=0.60(3)(2)$, with the second error due to
  an extrapolation uncertainty.}

\FullConference{The XXVII International Symposium on Lattice Field
  Theory - LAT2009\\  July 26--31, 2009\\
  Peking University, Beijing, China}

\begin{document}

\section{Introduction}

The strong coupling constant $\alpha_s=g^2/(4\pi)$ is one of the
$N_f+1$ input parameters of QCD and as such one of the fundamental
constants of nature. Its actual value
depends on both the renormalisation scheme (including the number of
active flavours) and the scale. Given a renormalisation scheme
$\RS$ the dependence on the scale $\mu$ is
controlled by the renormalisation group (RG) through 
\begin{equation}
  \mu^2\frac{d}{d\mu^2}\frac{\alpha^\RS_s(\mu^2)}{\pi} =
  \beta^\RS\left(\alpha^\RS_s \right)
  \; \stackrel{\alpha^\RS_s\to0}{\sim} \;  -
  \sum_{i\ge0}\beta^\RS_i\left(\frac{\alpha^\RS_s}{\pi} 
  \right)^{i+2},
  \label{eq:RGE}
\end{equation}
where $\beta^\RS$ is the beta function in that scheme. Solving
\Eq{eq:RGE}  
yields an exact relation between the scale-dependent coupling
$\alpha^\RS_s(\mu)$ and the RG-invariant, scale-independent
but renormalisation-scheme-dependent parameter $\Lambda_\RS$, 
defined via \mbox{$\ln
  \mu^2/\Lambda_\RS^2=\int d\alpha^\RS_s/\beta^\RS(\alpha)$}.
Once the Lambda parameter is known for one scheme, a one-loop 
calculation suffices to determine it in any other scheme. 

While recent precision determinations of 
$\alpha_s$ exist, based on either perturbative analyses of 
short-distance-sensitive lattice observables 
or sum rule analyses of hadronic $\tau$ decay data (for detailed discussions
and other relevant references, see \cite{Davies:2008sw, Maltman:2008bx}
and, e.g., \cite{Maltman:2008nf}, respectively),
residual uncertainties mean that
additional independent high-precision determinations remain of interest.
Some of us have recently introduced the minimal momentum subtraction
(MiniMOM, or $\MM$) scheme for QCD in covariant gauges
\cite{vonSmekal:2009ae}. An important advantage of this
scheme is that it allows the strong coupling to be
fixed solely through a determination of the gluon and ghost
propagators. In Landau gauge this scheme has been implicit in the
early studies of these propagators~\cite{vonSmekal:1997isvonSmekal:1997vx}.

The $\MM$ scheme is defined by combining $\MOM$ scheme
propagator renormalisation with the supplementary condition 
$\widetilde Z_1 = \widetilde Z_1^{\MSb}$ for the 
ghost-gluon vertex renormalisation constant~\cite{vonSmekal:2009ae}. 
With $Z$ and $G$
the respective gluon and ghost dressing functions, the $\MM$ coupling 
is then defined as
\cite{vonSmekal:1997isvonSmekal:1997vx}
\begin{equation} \alphaMM(p^2) \, = \, \frac{g^2}{4\pi} Z(p^2)
  G^2(p^2) \; .
  \label{alpha_minimom}
\end{equation}
The relation between $\alphaMM$ and $\alphaMS$ is known to four 
loops~\cite{vonSmekal:2009ae}.
Here we use $\alphaMM$ to determine $\LambdaMS^{(N_f)}$ (in
units of $r_0$) for \mbox{$N_f=0,2$} from continuum extrapolations of
the product of the bare lattice Landau gauge propagators, as first
proposed in \cite{Sternbeck:2007br}.\footnote{Our \mbox{4-loop}
expansion for 
$\beta^{\MM}(\alpha)$ was not worked out until January 2008. While the
\mbox{3-loop} version of this expansion in fact differs somewhat from
the 3-loop $\MOM h$ scheme result used in \cite{Sternbeck:2007br}, the
difference is small.}  The absence of vertex measurements in the method
allows for a significantly improved accuracy in the
lattice estimate for $\alphaMM$.

\section{Numerical setup}

The results below were obtained on both \mbox{$N_f=0$} and $2$\, $SU(3)$
gauge field configurations. The quenched configurations were
thermalised using the standard Wilson gauge action, with $\beta$
ranging from 6.0 to 8.5, applying standard update cycles of heatbath
and micro-canonical over-relaxation steps. The unquenched gauge field
configurations were provided by the QCDSF collaboration, who used the
same gauge action supplemented by \mbox{$N_f=2$} clover-improved
Wilson fermions at various values of the hopping-parameter~$\kappa$
(see \Tab{tab:para} for further details).  All gauge configurations
were fixed to lattice Landau gauge using an iterative gauge-fixing
algorithm. To guarantee high-precision the local violation of
transversality was not allowed to exceed \mbox{$\epsilon<10^{-10}$}
where, as usual,
$
\epsilon \equiv \max_x\, \Re\Tr\left[(\nabla_{\mu} A_{x\mu})(\nabla_{\mu}
    A_{x\mu})^{\dagger}\right]
$
and
\mbox{$ 
 A_{x\mu}\equiv
 \tfrac{1}{2iag }(U_{x\mu}-U_{x\mu}^{\dagger})|_{\mathrm{traceless}}
$}\,.

Gluon and ghost propagators were measured on these
gauge-fixed sets employing standard techniques and an acceleration for the
Faddeev-Popov-operator inversion (see
\cite{Sternbeck:2007br,Sternbeck:2005tk} for details). 
As verified numerically in \cite{Sternbeck:2005tk}, for
the range of momenta studied here the Gribov ambiguity is irrelevant.
Using well established values for $r_0/a$
\cite{Necco:2001xg,Guagnelli:2002ia,Gockeler:2005rv} to
bring the raw data on $\alphaMML$ for different $\beta$ (and $\kappa$)
onto the common scale $r_0^2 p^2$ (see \Tab{tab:para} for the $r_0/a$ values),
and with $g^2(a)=6/\beta(a)$ the bare coupling at the lattice
cutoff scale $a^{-1}$, $\alphaMM$ was then determined from the averaged
data for the bare lattice gluon and ghost propagator dressing functions,
$Z_L$ and $G_L$, via
\begin{equation}
  \label{eq:running_coup_latt}
  \alphaMM(p^2) = \alphaMML(p^2) + O(a^2) \qquad\text{with}\qquad
  \alphaMML(p^2) \equiv \frac{g^2(a)}{4\pi}\, Z_L(p^2,a^2)\,
  G^2_L(p^2,a^2) \,.
\end{equation}
To have the lattice tree-level structure correct the dressing functions
were extracted using $aq_\mu(p_\mu)=2\sin(ap_\mu/2)$, but $\alphaMM$
is considered versus $p^2$ with $ap_\mu = 2\pi k_\mu/L_\mu$ and
$k_\mu\in(-L_\mu/2,L_\mu/2]$.

\begin{table}
  \centering
  \caption{Parameters of gauge configurations. $\nicefrac{r_0}{a}$
    values are from
    \cite{Necco:2001xg,Guagnelli:2002ia,Gockeler:2005rv}; $a$[fm] 
    is for \mbox{$r_0=0.467\,\textrm{fm}$}.}  
  \label{tab:para}\small 
  \begin{tabular}{crrc@{\qquad}|@{\quad}cccrcc}
    \hline\hline
    $\beta$ & $N^4$  & \multicolumn{1}{c}{$r_0/a$} & $a$[fm] & $\beta$
    & $\kappa$ & $N_s^3\times N_t$ & \multicolumn{1}{c}{$r_0/a$} &
    $a$[fm] & $am_0$\\
    \hline
    6.0     & $32^4$ & 5.3677 & 0.087 
&  5.25 & 0.13575 & $24^3\times48$  & 5.532(40) & 0.0844 & 0.01414 \\
    6.2     & $32^4$   & 7.3829  & 0.060 
& 5.25 & 0.13600 & $24^3\times48$  & 5.732(64)
    & 0.0815  & 0.00737 \\
    6.4     & $32^4$   & 9.7415 & 0.048 
& 5.29 & 0.13590 & $24^3\times48$ 
    & 5.835(30) & 0.0800 &0.01456 \\
    6.6     & $32^4$   & 12.5955 & 0.037 
& 5.29 & 0.13620 & $24^3\times48$ 
    & 6.083(26) & 0.0768 &0.00646 \\
    6.9     & $48^4$   & 18.6757 & 0.025 
& 5.29 & 0.13632 &$32^3\times64$  
    & 6.153(62) & 0.0759 & 0.00323\\
    7.2     & $64^4$   & 27.11 & 0.017 
& 5.40 & 0.13610 & $24^3\times48$ 
    & 6.714(64) & 0.0696 &0.01575 \\ 
    7.5     & $64^4$   & 37.71 & 0.012 
& 5.40 & 0.13640 & $32^3\times64$  
    & 6.829(71) & 0.0690 &0.00767 \\
    8.5      & $48^4$    & 122.73   & 0.004 
& 5.40 & 0.13660 &$32^3\times64$ &  6.895(63)  & 0.0681 &0.00230 \\ 
           \hline\hline
  \end{tabular}\vspace{-2ex}
\end{table}

\section{Lattice data of the MiniMOM coupling}

\begin{figure*}
  \centering
    \includegraphics[width=0.8\linewidth]{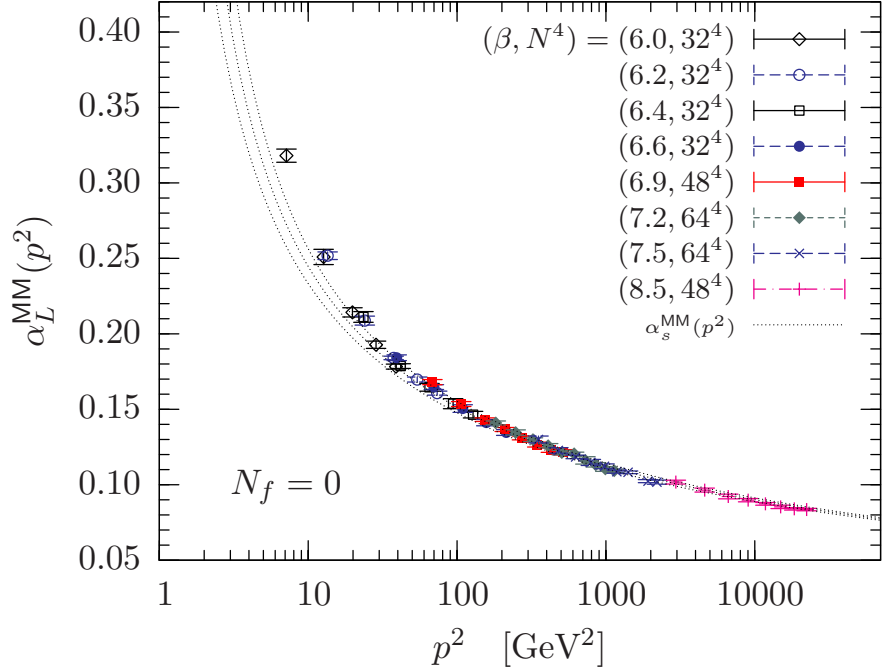}
    \\*[0.5cm]
     \includegraphics[width=0.8\linewidth]{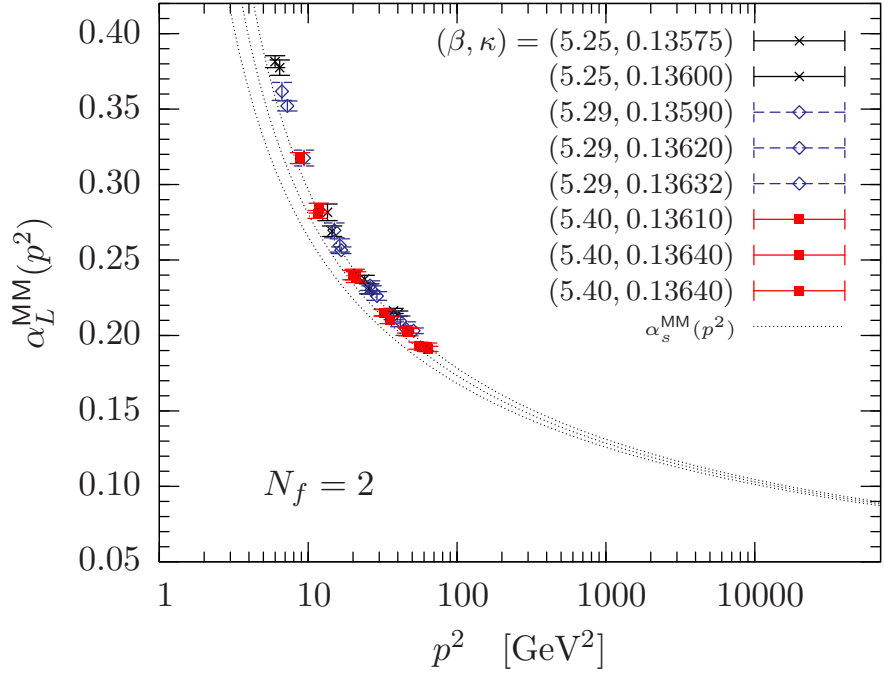}
  \caption{Lattice data for $\alphaMM(p^2)$ for \mbox{$N_f=0$}
    (top) and \mbox{$N_f=2$} (bottom) compared to the expected \mbox{4-loop}
    running of $\alphaMM(p^2)$ (lines) in the continuum limit
    \cite{vonSmekal:2009ae}. Note that these are not fits: The data
    is the raw lattice data for purely diagonal 
    momenta $ap_\mu=2\pi k_\mu/L_\mu$ with $1 < a^2p^2 < 10$
    brought onto a common scale employing established values for
    $r_0/a$ for these sets 
    \cite{Necco:2001xg,Guagnelli:2002ia,Gockeler:2005rv}.
    For illustration purposes, the 4-loop running has been fixed using
    \mbox{$r_0\Lambda_{\MSb}^{(0)}=0.60(5)$} and  
    \mbox{$r_0\Lambda_{\MSb}^{(2)}=0.62(4)(4)$} of the ALPHA
    collaboration \cite{Capitani:1998mq,DellaMorte:2004bc},
    and the overall momentum scale set
    using \mbox{$r_0=0.467\,\mathrm{fm}=2.367\mathrm{GeV}^{-1}$}.}  
  \label{fig:alphaMM_vs_4loop}
\end{figure*}

Our lattice data for $\alphaMML(p^2)$ is shown in
\Fig{fig:alphaMM_vs_4loop}, together with the expected 4-loop
continuum running obtained using, to
be specific, the ALPHA collaboration $\MSb$ values,
\mbox{$r_0\Lambda_{\MSb}^{(0)}=0.60(5)$,} and 
\mbox{$r_0\Lambda_{\MSb}^{(2)}=0.62(4)(4)$
}~\cite{Capitani:1998mq,DellaMorte:2004bc}, translated to
the MiniMOM scheme (see Table~4 of Ref.~\cite{vonSmekal:2009ae}
for the relevant values of $\LambdaMS/\LambdaMM$). In what follows, for the 
sake of illustration, we display momenta in physical units (rather than 
as $r^2_0p^2$), using 
\mbox{$r_0=0.467\,\mathrm{fm}=2.367\,\mathrm{GeV}^{-1}$}.
Since only ratios of momenta enter the calculation,
this choice does not affect the final result for $r_0\LambdaMS^{(N_f)}$.

From \Fig{fig:alphaMM_vs_4loop}, one sees that,
for both $N_f=0,\, 2$, scaling violations,
finite volume effects and hypercubic lattice
artefacts are nearly negligible, even though, for \mbox{$N_f=0$},
the lattice spacing varies over an order of magnitude. 
Small systematic deviations from continuum 4-loop running, however, 
become visible at higher resolution. Such deviations are negligible 
for purely diagonal lattice momenta satisfying
\mbox{$3<a^2p^2<6$}, but grow, \mbox{$\propto 1/(ap)^2$} (\mbox{$\propto
a^2p^2$}) to leading order, for momenta below (above) this interval.
\Fig{fig:alphaMM_vs_4loop} shows data for
diagonal momenta with \mbox{$1<a^2p^2<10$}.

Deviations from continuum 4-loop running are not unexpected. At small momenta,
they result from a mixture of (a)~finite volume effects, (b) the onset of
nonperturbative effects (condensates etc.) and
(c)~truncation errors in the perturbative expansion of the coupling,
while at large momenta they are due to (d)~scaling 
violations proportional to $a^2p^2$ and (e)~hypercubic lattice
artefacts proportional to higher-order invariants
$a^np^{[n]}=\sum_{\mu}a^np^{n}_\mu$ ($n=4,6,8$) of the isometry group 
$H(4)$. The latter, in particular, would become pronounced for larger 
momenta without suitable corrections, which can be performed either 
using the H4 method or by imposing so-called cylinder cuts on the data 
(for the former approach, see, e.g., \cite{deSoto:2007ht}, for the latter
\cite{Leinweber:1998uu}). The cylinder cut approach, though less 
sophisticated, is nonetheless effective and robust. We chose
a combination of the two methods to keep the statistical
noise to a minimum. To be specific, we consider data 
only for \emph{purely diagonal} lattice momenta, for which hypercubic
lattice artefacts are known to be 
smallest, and correct for the remaining (rather small) artefacts 
through a fit of this data to a hypercubic Taylor expansion of
the (lattice) $\MM$ coupling which, to leading order, has the form
\begin{equation}
  \label{eq:running_coup_latt_plus_corrections}
  \alphaMML(p^2) = \alphaMM(p^2)\left(1 + c_1\cdot a^2p^2 + c_2 \cdot
  a^4p^4 + \ldots\right)
\end{equation}
\begin{floatingfigure}[r] 
   \centering
   \parbox{6.8cm}{%
    \includegraphics[width=7.4cm]{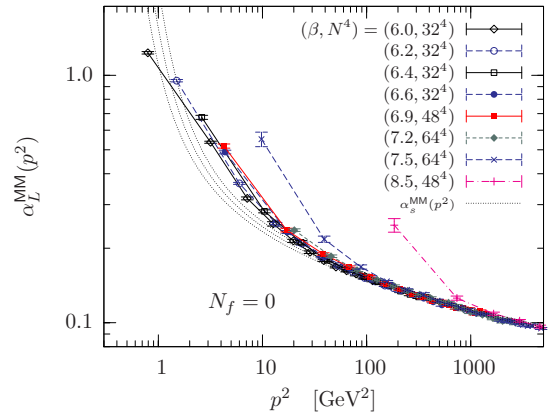}}
    \caption{\mbox{$N_f=0$} data for $\alphaMML(p^2)$ as shown in
      \protect\Fig{fig:alphaMM_vs_4loop} but here for $0<a^2p^2<23$
      to illustrate finite volume effects at small momenta
      which are more pronounced here than the lattice artefacts at
      large.}
    \label{fig:alphaMM_pp_nf0_raw_small_pp}
\end{floatingfigure}
\vspace{-0.7ex}
\noindent where the $c_i$ are constants (see also
\cite{deSoto:2007ht,DiRenzo:2006wd,Boucaud:2008gnDesoto:2009aw}). 
For the classic H4 method one would have to fit the $c_i$'s from
extrapolations of the $\alphaMML(p^2)$ data at different lattice
momenta belonging to different $H(4)$ orbits but the same
$a^2p^2$. Not only are these extrapolations susceptible
to statistical artefacts for insufficiently large lattice sizes,
but the Faddeev-Popov operator has to be inverted 
using point sources to get data for all momenta,
introducing larger statistical fluctuations into the ghost
propagator (and hence the coupling) at large momenta. Our
approach allows us to use instead plane-wave sources for our inversions 
(see \cite{Sternbeck:2005tk} for details) and fully exploits
the translation invariance of the lattice, thus drastically reducing
statistical noise in the coupling. The results below
bear out the reasonableness of the approach, providing
an excellent description of the data
for \mbox{$3<a^2p^2<30$} (or \mbox{$3<a^2p^2<12$} for
$c_2\equiv 0$) and giving stable results for all fit parameters.

The deviations at small momenta, illustrated in
\Fig{fig:alphaMM_pp_nf0_raw_small_pp}, are more severe and, in our
opinion, not yet fully under control.  They start to become visible
for $a^2p^2<1$ and appear to be a mixture of finite volume and
nonperturbative effects (the latter expected to set in at smaller
$\beta$). The effect is such that data points at fixed physical
momenta decrease as the physical volume increases. Currently,
additional simulations at different $\beta$ but fixed physical volume,
are being performed to help bring these low-momentum artefacts under
better control. For now, we exclude data with $a^2p^2<3$ to stay well
clear of the region where such effects become evident.

\section{Fitting the data}

Our fitting procedure works as follows. Each of the data sets is
fitted separately to the
Ansatz~\eqref{eq:running_coup_latt_plus_corrections}, where
the 4-loop perturbative running form is
used for $\alphaMM(p^2)$ and the remaining terms 
correct for the leading lattice artefacts at larger $a^2p^2$. 
All fits are performed using the fixed 
fitting window \mbox{$3<a^2p^2<12$}. (The fits have been checked to be
quite robust to small changes to the lower and upper bounds of this
window.) We also find that $c_2$ can be set to $0$ with little
effect on $c_1$. The 4-loop running of $\alphaMM (p^2)$ is
generated from that of $\alphaMS (p^2)$ using the 4-loop
relation between the couplings given in
Eqs.~(14-15) of Ref.~\cite{vonSmekal:2009ae}. (Further, more
specific details will be provided in an upcoming publication.)

With $c_2$ set to zero, our fit parameters reduce to $c_1$, 
the constant of the leading lattice correction at larger momenta, and
$\alphaMSref$, the $\MSb$ coupling at an
arbitrary reference scale $p^2_{\mathrm{ref}}$ (to be specific, we take
$p^2_{\mathrm{ref}}=70\,\textrm{GeV}^2$ and
$r_0=0.467\,\textrm{fm}$). $\LambdaMS$ could, of course, be
used in place of $\alphaMSref$ as a fit parameter; we
expect our fits to be more stable with the latter choice.
$\Lambda_{\MSb}$ in any case follows from $\alphaMSref$ using
the standard relation \cite{Chetyrkin:1997iv}
\begin{displaymath}
  \label{con}
  \ln\frac{\mu^2}{\Lambda^2}=\int\frac{da}{\beta(a)}
  =\frac{1}{\beta_0}\left[\frac{1}{a}+b_1\ln a+(b_2-b_1^2)a
    + \left(\frac{b_3}{2}-b_1b_2+\frac{b_1^3}{2}\right)a^2\right]+C
\end{displaymath}
where $a(\mu)\equiv\alpha_s(\mu)/\pi$, 
$\beta^{\MSb}_0,\ldots,\beta^{\MSb}_3$ are the $\MSb$ scheme 
$\beta$-function coefficients, $b_i=\beta_i/\beta_0$ 
and $C=(b_1/\beta_0)\ln\beta_0$.

\medskip

\begin{figure}[t]
  \centering
  \mbox{\includegraphics[width=0.5\linewidth]{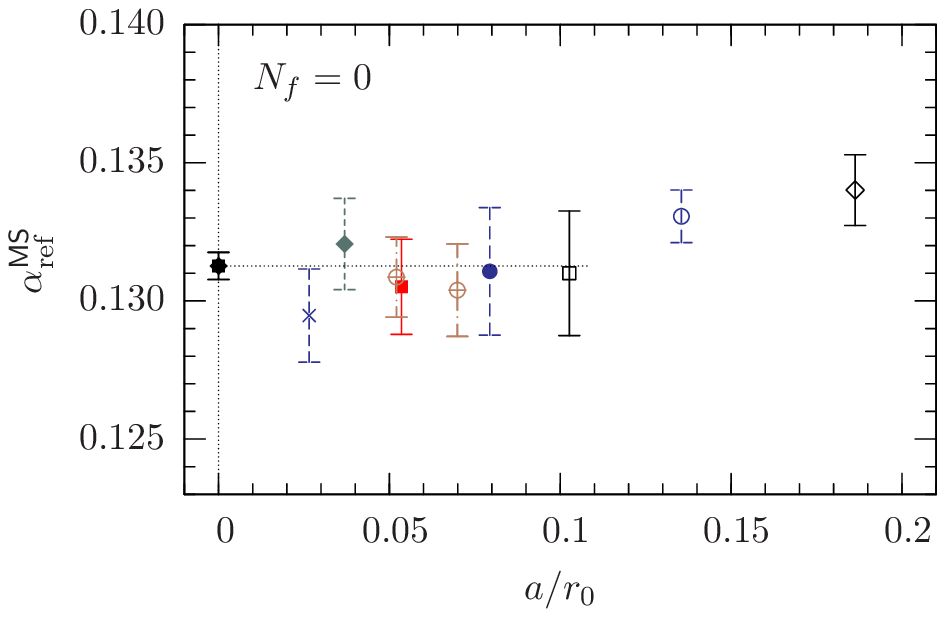}\,
    \includegraphics[width=0.5\linewidth]{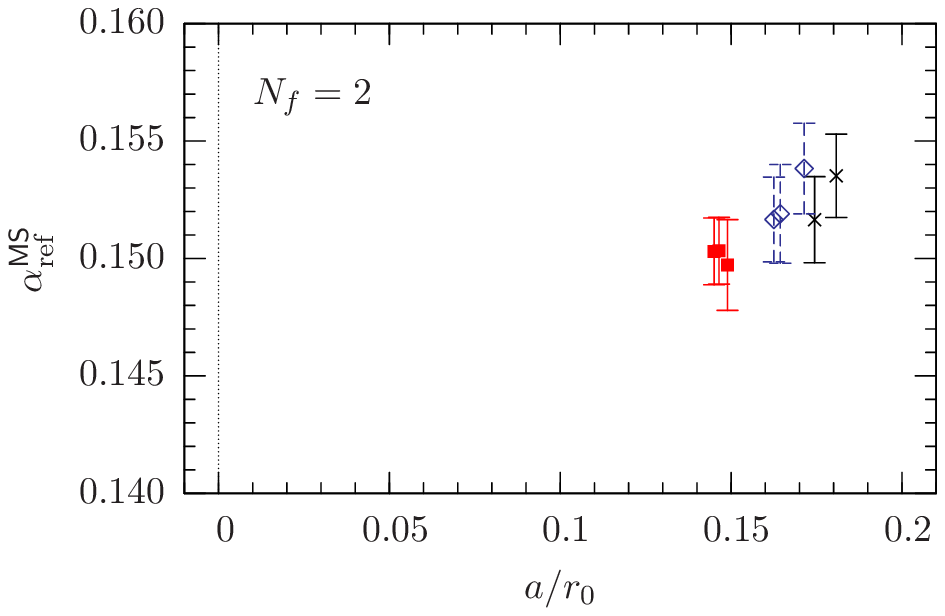}}
  \caption{Values for $\alpha^{\MSb}_{\mathrm{ref}}$ obtained from
    fits to the data for different $a/r_0$ using Ansatz
    \protect\eqref{eq:running_coup_latt_plus_corrections}; left for
    $N_f=0$ and right   
    for $N_f=2$ and different bare quark masses $am_0$. See
    text and \protect\Tab{tab:para} for further details. Symbols
    refer to the same $\beta$, $N^4$ (and $\kappa$) as in
    \protect\Fig{fig:alphaMM_vs_4loop}. The dotted 
    line and very left `star' (left panel) are a fit of the
    plateau of the three points (filled symbols), 
    calculated in equal physical volumes
    $V_1\approx(1.2\,\mathrm{fm})^4$. Note that there we have also
    included two very recent points (brown crossed circles) from
    simulations at $\beta=6.7$ and $6.92$
    and equal physical volume, $V_2\approx(1.6\,\mathrm{fm})^4$.}
  \label{fig:alphaMS_ref_nf_a2}
\end{figure}

\begin{figure}[t]
  \centering
  \mbox{\includegraphics[width=0.5\linewidth]{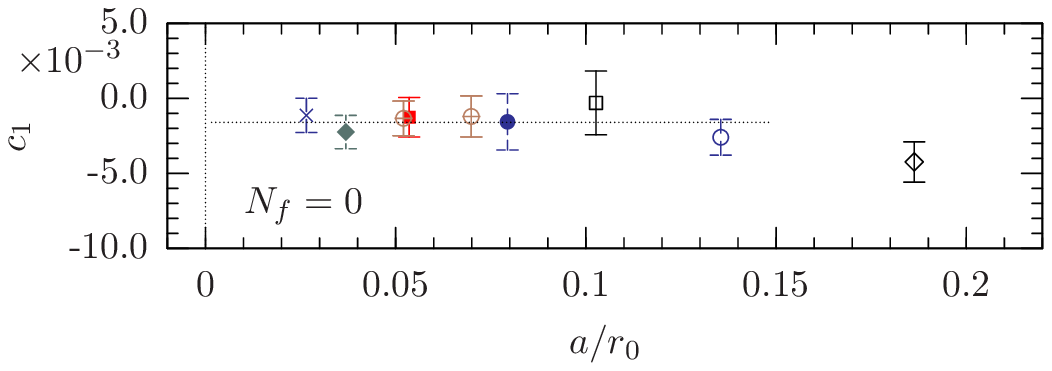}\,
    \includegraphics[width=0.5\linewidth]{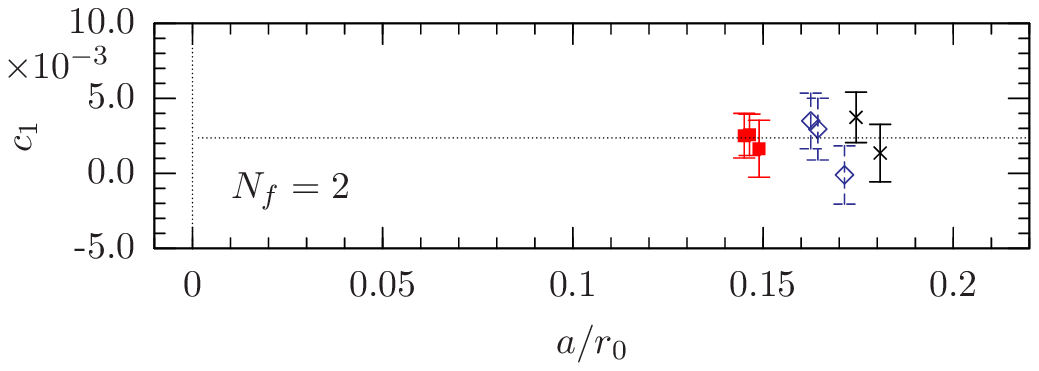}}
  \caption{Coefficient $c_1$ of the leading lattice correction
    (see \protect\Eq{eq:running_coup_latt_plus_corrections}) for different
    $a/r_0$, obtained from the same fits as
    $\alpha^{\MSb}_{\mathrm{ref}}$ shown in
    \protect\Fig{fig:alphaMS_ref_nf_a2}. Symbols are the same as in
    Figs.~\protect\ref{fig:alphaMM_vs_4loop} and
    \protect\ref{fig:alphaMS_ref_nf_a2}. Dotted lines mark the 
    average $c_1$ including data left for $a/r_0<0.15$ and right for
    $a/r_0<0.2$.}
  \label{fig:c_nf_a2}
\end{figure}

Fitted values for $\alphaMSref$ and $c_1$ as a function of $a/r_0$ 
(and, for $N_f=2$, also for different bare quark masses)
are shown in Figs.~\ref{fig:alphaMS_ref_nf_a2} and \ref{fig:c_nf_a2}. 
Note that if the data showed perfect scaling, all 
fits would give the same value (within errors) for $\alphaMSref$,
independent of the lattice spacing, and the fitted $O(a^2)$ 
corrections ($c_1$) would turn out to be zero. As expected,
$c_1$, though small, is not zero (see \Fig{fig:c_nf_a2}). The 
long plateau for $c_1$ as a function of $a/r_0$, however,
suggests that our Ansatz for describing the leading lattice
corrections is a reliable one. This conclusion is also supported
by the results for $\alphaMSref$. For \mbox{$N_f=0$}, for example, 
$\alphaMSref$ starts to plateau around $a/r_0=0.1$,
i.e., for $\beta\ge6.4$ (see
\Fig{fig:alphaMS_ref_nf_a2}). Note that the fit quality is
significantly degraded if no correction term is included.


For \mbox{$N_f=2$} we observe only small deviations in
$\alphaMSref$ on changing the bare quark mass (see 
\Fig{fig:alphaMS_ref_nf_a2}, right panel). Changing $a/r_0$ leads to more
significant shifts. This is almost certainly due to the relatively large 
$a/r_0$ employed, and the fact that the lattice data available for
$\alphaMM$ come from momenta where deviations from the 4-loop running
of $\alphaMM$ have already set in 
(see ~\Fig{fig:alphaMM_vs_4loop}, bottom panel). The overall
picture, however, appears similar to that for $N_f=0$, and the fitted
$\alphaMSref$ for the $\beta=5.40$ configurations are already 
close to what is expected, for example, from  \cite{DellaMorte:2004bc}. 
In fact, $r_0\Lambda_{\MSb}^{(2)}=0.62(4)(4)$ corresponds to  
$\alphaMSref=0.150(3)$ in the right panel of
\Fig{fig:alphaMS_ref_nf_a2}.

\medskip

A continuum extrapolation in our approach corresponds
to fitting data plateaus visible at small enough
$a/r_0$. While it is far from clear that a plateau
has been reached for the $N_f=2$ data in 
\Fig{fig:alphaMS_ref_nf_a2} the $N_f=0$ $\alphaMSref$ results
level off quite nicely at smaller $a/r_0$. 
Fitting the three points having the same physical volume, i.e.,
those with $(\beta,N)=\left\{(6.6,32), (6.9,48), (7.2,64)\right\}$,
yields $\alphaMSref=0.131(1)$. This corresponds to
$r_0\Lambda_{\MSb}^{(0)}=0.62(1)$ which agrees well with values
from the literature. Assuming that the relative decrease of
$\alphaMSref$ from $a/r_0=0.15$ to $\sim 0.1$ will be the same for $N_f=0$
and $2$ (about 1.5\%), one would expect the $N_f=2$ data to level
off at smaller $a/r_0$ around $\alphaMSref=0.148(2)$. This would
correspond to $r_0\Lambda_{\MSb}^{(2)}=0.59(3)$, also in good agreement
with existing values in the literature. Additional data for
$N_f=2$ at smaller $a/r_0$ would allow us to make further progress,
but adequate gauge configurations are unfortunately not yet available.
For now we take the average $r_0\Lambda_{\MSb}^{(2)}=0.60(3)(2)$ with
the second error due to the uncertainty in the continuum extrapolation.

\section{Conclusions}

In this paper, we have taken advantage of the recently introduced
$\MM$ scheme for QCD in covariant gauges~\cite{vonSmekal:2009ae} to
perform a determination of the QCD Lambda parameter for $N_f=0,2$.
The scheme allows the strong coupling constant, and hence
$\LambdaMS$, to be determined from measurements of ghost and gluon
two-point functions on the lattice. The restriction to measured
two-point functions, and the fact that the relation between
the $\MM$ and $\MSb$ couplings is known to 4-loop order,
allows for a high precision determination, with reliable
error estimates.

Our results to date are restricted to $N_f=0,2$, and must
be extended to $N_f=2+1$ in order to reach the desired
goal of estimating $\alphaMS(M_Z)$. Fortunately,
over the last few years, the number of available $N_f=2+1$ gauge 
configurations has increased significantly, and continues to increase.
The $N_f=0,2$ results presented
above, which yield the (still preliminary) results
\begin{equation}
  \label{eq:current_estimates_for_r0Lambda}
  r_0\Lambda_{\MSb}^{(0)} = 0.62(1)\, \qquad\textrm{and}\qquad
  r_0\Lambda_{\MSb}^{(2)} = 0.60(3)(2) ,
\end{equation}
in agreement with other studies (e.g.,
\cite{Gockeler:2005rv,Capitani:1998mq,DellaMorte:2004bc,
Boucaud:2008gnDesoto:2009aw}), demonstrate the reliability and
accuracy of our method, and thus pave the way for future $N_f=2+1$
analyses.  The analysis also provides valuable information on how to
bring lattice artefacts under control.

A positive feature of the current study is that lattice artefacts are 
found to be almost negligible if one restricts the analysis to 
strictly diagonal lattice
momenta satisfying \mbox{$3<a^2p^2<6$}. For larger momenta,
hypercubic lattice artefacts become visible. For 
$a^2p^2<12$ they grow like $c_1\,\alphaMM(p^2)\, a^2p^2$, with
$c_1=-0.00016(3)$ for $N_f=0$, and $c_1=0.00023(4)$ for $N_f=2$, and thus
can be corrected for quite precisely. It is anticipated
that the precision could be improved further if data from lattice
perturbation theory was available (see, e.g,
\cite{DiRenzo:2009niDiRenzo:2009ei} for steps in this direction).

For smaller momenta, finite volume effects are present, in particular for
$a^2p^2<1$. These effects are not yet fully under control and new
calculations at different $\beta$ but fixed volumes are under way to
rectify this situation. It is for this reason that we have not yet
investigated condensate effects, which are expected to be relevant for
lower momenta (see, for example, \cite{Boucaud:2008gnDesoto:2009aw}).

Note that the $\MM$ coupling could also
be employed to determine the lattice spacing dependence $a(\beta)$
via $a\LambdaMS(\beta)$. With this information,
and the ratios of the $\Lambda_{\MSb}^{(N_f)}$ for $N_f=0, 2$ and 3, a
well-established experimental value for $\alphaMS(M_Z)$
could be used to fix the physical scale of $a$ for different $N_f$
(rather than using $r_0$). This again  would require a good
understanding of all lattice artefacts since $\alphaMM$ data for
different lattice spacings would have to be brought to a common
scale via a matching procedure, starting in the perturbative region.

\bigskip

\begin{spacing}{0.9}
{\small
This research was supported by the Australian Research Council. A.S.\
is also supported by the \mbox{Sfb/Tr-55}, K.M.\ by the Natural
Sciences and Engineering Council of Canada, and L.v.S by the
Helmholtz International Center for FAIR within the LOEWE program of
the State of Hessen, Germany. K.M.\ also acknowledges the hospitality
of the CSSM at the University of Adelaide. Grants of time on the
computing facilities of the HLRN-Verbund (Germany) and eResearchSA
(Australia) are acknowledged.}
\end{spacing}


\providecommand{\href}[2]{#2}\begingroup\raggedright
\endgroup

\end{document}